\documentclass[superscriptaddress,showkeys,twocolumn,nofootinbib,longbibliography]{revtex4-1}

\usepackage{amsmath}
\usepackage{amssymb}
\usepackage{graphicx}
\usepackage{epsf}
\usepackage{slashed}
\usepackage{enumitem}
\usepackage[czech,english]{babel}
\usepackage[cp1250]{inputenc}
\usepackage{hyperref}
\usepackage{xcolor}
\usepackage{mathtools}

\usepackage[only,llbracket,rrbracket,llparenthesis,rrparenthesis]{stmaryrd} % for comparison with 4 similar parenthesis symbols
\usepackage{accsupp} % for ensuring the right Unicode codepoint upon pasting

\newcommand{\diag}{\mathop{\rm diag}\nolimits}% diagonal matri
\newcommand{\e}{\ensuremath{\mathrm{e}}}% Euler number
\newcommand{\eL}{\mathcal{L}}% Lagrangian
\renewcommand{\d}{\ensuremath{\mathrm{d}}}% differential
\newcommand{\threevector}[1]{\boldsymbol{#1}}% three-vector -- jeste zvazit...
\newcommand{\qm}[1]{``#1''} % quotation marks
\newcommand{\sgn}{\mathop{\rm sgn}\nolimits}
\newcommand{\Ei}{\mathop{\rm Ei}\nolimits}
\newcommand{\EM}{\ensuremath{\gamma}}

\newcommand{\innerdot}{\ensuremath{\!\cdot\!}}% dot product

\newcommand{\group}[3]{#1(#2)_{\mathrm{#3}}}

\usepackage{bbm} %\mathbbm{1}, \mathbbmss{1}, \mathbbmtt{1}

\usepackage[nodayofweek]{datetime}
\newdateformat{mydate}{\twodigit{\THEDAY}{ }\shortmonthname[\THEMONTH], \THEYEAR}
\usepackage[normalem]{ulem}
\begin{document}

\title{Magnetic monopoles with an internal degree of freedom}

\newcommand{\affiliacePraha}{Institute of Experimental and Applied Physics, \\ Czech Technical University in Prague, Husova~240/5, 110~00 Prague~1, Czech Republic}

\newcommand{\affiliaceOpava}{Research Centre for Theoretical Physics and Astrophysics, Institute of Physics, \\ Silesian University in Opava, Bezru\v{c}ovo n\'{a}m\v{e}st\'{\i}~1150/13, 746~01 Opava, Czech Republic}

\author{Petr Bene\v{s}}
\email{petr.benes@utef.cvut.cz}
\affiliation{\affiliacePraha}

\author{Filip Blaschke}
\email{filip.blaschke@fpf.slu.cz}
\affiliation{\affiliacePraha}
\affiliation{\affiliaceOpava}

\begin{abstract}
We consider a class of spontaneously broken $\group{SU}{2}{}$ gauge theories with adjoint scalar and look for exact magnetic monopole solutions in the Bogomol'nyi--Prasad--Sommerfield (BPS) limit. We find that some of the resulting solutions exhibit a new internal degree of freedom (a moduli space parameter) that controls the energy density profile of the monopole while keeping the total energy (mass) constant.
%\\ \remark{Compiled: \mydate\today, {\currenttime}, version V6}
\end{abstract}


\maketitle

%\tableofcontents

\section{Introduction}
\label{intro}

Although as-yet unobserved, the magnetic monopoles are undoubtedly among the most intriguing particles, at least from the conceptual point of view. While all other particles (both observed and hypothetical) are given by (quantized) linear perturbations of the fields above their minimum values, the magnetic monopoles are given non-perturbatively, by full (classical) solutions of the non-linear equations of motion.

The father of modern theory of magnetic monopoles is undoubtedly Dirac, who devised the celebrated quantization condition \cite{Dirac:1931kp}: All magnetic and electric charges in universe must be related as $q_{\mathrm{e}} q_{\mathrm{m}} = 2\pi n$, where $n$ is an integer.

The story continued in 70' with the Georgi--Glashow model \cite{Georgi:1972cj},
\begin{eqnarray}
\label{lagrangianGG}
\eL &=& 
\frac{1}{2} (D^\mu \threevector{\phi})^2
- \frac{1}{4g^2} (\threevector{F}^{\mu\nu})^2
- V(\threevector{\phi}^2)
\,,
\end{eqnarray}
an $\group{SU}{2}{}$ gauge theory with adjoint scalar spontaneously broken down to \qm{electromagnetic} $\group{U}{1}{}$. (We are using boldface to denote an adjoint vector.) This theory contains in its spectrum a massless neutral vector boson (the photon), two charged massive vector bosons (the $W^{\pm}$), and a neutral massive scalar (the Higgs boson). As such, the model \eqref{lagrangianGG} was once, before the discovery of neutral currents, considered a candidate for a theory of electroweak interactions.

Surprisingly, it was found in 1974, independently by 't~Hooft \cite{tHooft:1974kcl} and Polyakov \cite{Polyakov:1974ek}, that in addition to the mentioned \emph{perturbative} spectrum, there exists also its \emph{non-perturbative} counterpart -- the magnetic monopole. Mathematically, it is described as a topological soliton which, loosely speaking, exploits the fact that embedding of the unbroken $U(1)$ in the parent $\group{SU}{2}{}$ can be different in different space directions. This gives rise to a topological defect with particle-like properties -- the magnetic monopole.

Remarkably, the \emph{existence} of monopoles is ensured solely by the given topology of the system -- the interplay between the manifold of spatial infinities and the vacuum manifold. On the other hand, the particular form of interactions between the gauge bosons and scalars doesn't play any r\^{o}le in this regard. Nevertheless, the interactions do influence the \emph{properties} of the monopole -- its \qm{shape} (that is, the energy density profile) and, in general, also its mass.

Therefore, it is a natural idea to modify the interactions in \eqref{lagrangianGG} while keeping the topology intact, and observe how the monopole solution is affected by these modifications. The goal is twofold: To find new analytic monopole solutions and to identify models that have monopoles with genuinely new properties compared with the Georgi--Glashow model as a benchmark.

In order to go beyond \eqref{lagrangianGG} without adding new fields, one necessarily has to give up the requirement of renormalizability and introduce non-minimal interactions. However, since the space of all such non-renormalizable extensions is huge, a guiding principle is needed. To that end, we first note the trivial fact that the Lagrangian \eqref{lagrangianGG} contains no more than two derivatives of both the gauge fields, $\threevector{F}^{\mu\nu}$, and of the scalars, $D^\mu \threevector{\phi}$. So the idea suggests itself: To consider \emph{the most general} model with this property.\footnote{This limitation is advantageous also for another reason: These theories are safe from the instabilities (Ostrogradski ghosts) that can occur if more than two (time) derivatives are present since in such a case the Hamiltonian is not bounded from below. As such, the models we wish to deal with are the most general classical theories with \qm{guaranteed good behavior}.}

Thus, a na\"{\i}ve conclusion might be that a Lagrangian of such a theory would be, besides the potential term, simply a linear combination of the usual kinetic terms,
\begin{equation}
\label{termsKinOld}
(D^\mu \threevector{\phi})^2 \,,
\hspace{10mm}
(\threevector{F}^{\mu\nu})^2 \,,
\end{equation}
just like \eqref{lagrangianGG}, but this time with the coefficients being some gauge-invariant functions of $\threevector{\phi}$.

Indeed, the resulting theories with these so-called \qm{non-canonical} kinetic terms have been extensively studied in the literature already for a while. The first time such a theory was considered was in \cite{Casana:2012un}, where the model was formulated and numerical monopole solutions (in the Bogomol'nyi--Prasad--Sommerfield (BPS) limit), whose existence was later proved more rigorously in \cite{Zhang:2018ohp}, were found. Later, in \cite{Casana:2013lna, Ramadhan:2015qku} analytical solutions were found as well, although in a special way: First an analytical solution for $\threevector{\phi}$ was postulated and only then a solution for $\threevector{A}^\mu$ and the corresponding model were calculated. In \cite{Bazeia:2013nma} method of finding a new solution by \qm{deforming} an already known one was introduced. The quest for new monopole solutions, both analytical and numerical, in this class of models continued and resulted in monopoles with interesting new properties: In \cite{Bazeia:2018fhg} the so-called \qm{hollow monopoles} (having vanishing energy density in their center) were first reported, though only numerically, while in \cite{Bazeia:2018eta} monopoles with \qm{internal structure} (energy density having more complicated profile) were found (within models containing an additional scalar singlet). Finally, a generalized concept of self-duality in similar models was studied in \cite{Ferreira:2021uhk}.

However, linear combinations of only \eqref{termsKinOld} don't exhaust all possibilities. In fact, there are also two other terms,
\begin{equation}
\label{termsKinNew}
(\threevector{\phi} \innerdot D^\mu \threevector{\phi})^2 \,,
\hspace{10mm}
(\threevector{\phi} \innerdot \threevector{F}^{\mu\nu})^2 \,,
\end{equation}
whose algebraic independence on \eqref{termsKinOld} is due to the non-trivial representation of $\threevector{\phi}$ under the non-Abelian gauge group.

The new terms \eqref{termsKinNew} have been first introduced in our recent work \cite{Benes:2023nsr}. There we have formulated the full model, constructed the BPS limit, and found a whole new class of exactly solvable models that directly generalize the standard Prasad and Sommerfield solution \cite{Prasad:1975kr} of the 't~Hooft--Polyakov monopole. We found analytical solutions for both the hollow monopole, first discovered numerically in \cite{Bazeia:2018fhg}, and for the monopoles with internal structure, similar to those reported in \cite{Bazeia:2018eta}, but without resorting to the inclusion of additional fields.

In the present work\footnote{Some of the content presented here has been already partially presented in the conference proceeding \cite{Benes:2024evt}; in this paper, we are providing the full and more detailed treatment.} we continue in this program. Within the same model as in \cite{Benes:2023nsr} we devise \qm{yet another} class of analytic BPS monopole solutions. Remarkable enough, however, these new solutions happen to possess an interesting and completely unexpected new feature: They depend on a parameter (dubbed $\xi$) that is not present in the Lagrangian! This parameter can vary smoothly in a certain range without changing the total energy, the monopole mass.  It does, however, influence the energy density profile. Thus, this parameter is a measurable, physical quantity and can be interpreted as a kind of internal degree of freedom of the monopole -- a zero mode for the BPS monopoles in these special models.

The paper is organized as follows. First, in Sec.~\ref{theorynonbps} we mostly recapitulate some of the results from \cite{Benes:2023nsr}, in order to make the present text self-contained. The core of the paper is in Sec.~\ref{InvertibleCaseGeneral} where we derive the new class of analytic solutions and analyze some of its properties, most notably the presence of the aforementioned new parameter $\xi$. In Sec.~\ref{examples} we provide some particular examples. Finally, in Sec.~\ref{conclusions} summarize and discuss our results.

We are using the conventions $g_{\mu\nu} = \diag(+,-,-,-)$, $\varepsilon^{123} = +1$ and $c = \hbar = 1$.

\section{The stage}
\label{theorynonbps}

\subsection{The general (non-BPS) Lagrangian}

\newcommand{\HtoH}{\ensuremath{\alpha}}

As argued above, we consider the most general spontaneously broken $\group{SU}{2}{}$ gauge theory with an adjoint scalar that is quadratic in derivatives. As such, it must be a linear combination of both the \qm{usual} kinetic terms $(\threevector{F}^{\mu\nu})^2$, $(D^\mu \threevector{\phi})^2$, and the \qm{new} terms $(\threevector{\phi} \innerdot \threevector{F}^{\mu\nu})^2$, $(\threevector{\phi} \innerdot D^\mu \threevector{\phi})^2$, plus the potential term. Moreover, the coefficients of this linear combination are allowed to be arbitrary smooth gauge-invariant functions of $\threevector{\phi}$.

Although the final form of the Lagrangian will be eventually written in a different basis, at this point the most convenient way is to write it in terms of projector-like structures and with negative powers of $\threevector{\phi}^2$ explicitly factorized out, i.e.,
\begin{eqnarray}
\label{lagrangian}
\eL &=& 
\frac{v^2}{2} \bigg[
f_1^2 \bigg(
\frac{(D^\mu \threevector{\phi})^2}{\threevector{\phi}^2}
- \frac{(\threevector{\phi} \innerdot D^\mu \threevector{\phi})^2}{\threevector{\phi}^4}
\bigg)
+
f_3^2 \frac{(\threevector{\phi} \innerdot D^\mu \threevector{\phi})^2}{\threevector{\phi}^4}
\bigg]
\nonumber \\ && {}
- \frac{1}{4g^2} \bigg[
f_2^2 \bigg(
(\threevector{F}^{\mu\nu})^2
-\frac{(\threevector{\phi} \innerdot \threevector{F}^{\mu\nu} )^2}{\threevector{\phi}^2}
\bigg)
+
f_4^2 \frac{(\threevector{\phi} \innerdot \threevector{F}^{\mu\nu})^2}{\threevector{\phi}^2}
\bigg]
\nonumber \\ && {}
- V(\threevector{\phi}^2)
\,,
\end{eqnarray}
where
\begin{eqnarray}
D^\mu \threevector{\phi} &=& \partial^\mu \threevector{\phi} + \threevector{A}^\mu \times \threevector{\phi} \,,
\\
\threevector{F}^{\mu\nu} &=& \partial^\mu \threevector{A}^\nu - \partial^\nu \threevector{A}^\mu + \threevector{A}^\mu \times \threevector{A}^\nu \,,
\end{eqnarray}
and $V$ is a potential that need not to be specified besides the fact that it has a minimum at $\threevector{\phi}^2=v^2$.

Importantly, the \emph{form-functions} $f_i^2 \geq 0$ are dimensionless gauge-invariant functions of $\threevector{\phi}$. Na\"{\i}vely, one might expect them to be functions of $\threevector{\phi}^2/v^2$. However, we can do slightly better. The scalar triplet can be decomposed as
\begin{eqnarray}
\label{definitionvHn}
\threevector{\phi} &=& v H \threevector{n} \,,
\end{eqnarray}
where the isovector $\threevector{n}$ is normalized as $\threevector{n}^2 = 1$ and the form factor $H$ is a dimensionless gauge-invariant scalar function. Thus, we will assume $f_i^2$ to be functions of $H$, not of $H^2 = \threevector{\phi}^2/v^2$:
\begin{eqnarray}
f_i^2 &\equiv& f_i^2(H) \,.
\hspace{10mm}
(i=1,2,3,4)
\end{eqnarray}
%Finally, 
Moreover, 
for convenience and without loss of generality, we will assume the normalization
\begin{eqnarray}
\label{finormalization}
f_1^2(1) \ = \ f_2^2(1) &=& 1 \,.
\end{eqnarray}

Finally, let us stress that the negative powers of $\threevector{\phi}$ in \eqref{lagrangian} are present only for convenience and are by no means indispensable, as they can be always absorbed into redefinition of the form-functions $f_i^2$.

\subsection{Form-invariance and redundancy}
\label{forminvariance}

The Lagrangian  \eqref{lagrangian} defines a class of models labeled by five functions $f_i^2$ and $V$. However, it turns out that this description is redundant in the sense that seemingly different Lagrangians (i.e., with different $f_i^2$, $V$) can be physically equivalent.

To see this, let us consider a transformation (field redefinition) of the original scalar field $\threevector{\phi} = vH\threevector{n}$ to a new field $\tilde{\threevector{\phi}} = v\tilde{H}\threevector{n}$, where $H = \HtoH(\tilde{H})$, with $\HtoH(\tilde{H})$ being an arbitrary invertible and differentiable function. Under this transformation, the original Lagrangian \eqref{lagrangian} transforms into a new Lagrangian \emph{of the same form}, but with $\tilde V(\tilde{\threevector{\phi}}^2) = V(\threevector{\phi}^2)$ and
\begin{eqnarray}
\label{transformationhi}
\tilde f_i^2(\tilde H) &=&
\begin{cases}
f_i^2\big(\HtoH(\tilde H)\big) \,, & (i=1,2,4) \\[2pt]
\displaystyle
f_i^2\big(\HtoH(\tilde H)\big) \bigg(\tilde{H}\frac{\HtoH^\prime(\tilde H)}{\HtoH(\tilde H)}\bigg)^2 \,. & (i=3)
\end{cases}
\hspace{10mm}
\end{eqnarray}
In other words, our class of models is \qm{form-invariant} under rescaling of $H$.

Since the field redefinition cannot change the physics, the Lagrangians with two different sets of defining functions, $f_i^2$, $V$ and $\tilde f_i^2$, $\tilde V$, are physically equivalent.

\subsection{The BPS limit}
\label{theorybps}

So far we have been using the parameterization of the Lagrangian \eqref{lagrangian} in terms of the form-functions $f_i^2$. However, to discuss the BPS limit it turns out that there is a more convenient set of form-functions, $F_i$, that are related to the original ones, $f_i^2$, as
\begin{subequations}
\begin{align}
f_1^2 &=
F_1 F_2 \,,
&
f_2^2 &=
F_1/F_2 \,,
\\
f_3^2 &=
H^2 F_3^\prime F_4^\prime \,,
&
f_4^2 &=
F_3^\prime/F_4^\prime \,,
\end{align}
\end{subequations}
or inversely as
\begin{subequations}
\label{Finf}
\begin{align}
F_1
&= f_1 f_2 \,,
&
F_2
&= f_1/f_2 \,,
\\
H F_3^\prime
&= f_3 f_4 \,,
&
H F_4^\prime
&= f_3/f_4 \,.
\end{align}
\end{subequations}
The new form-functions are normalized as
\begin{equation}
\label{Finormalization}
\big|F_i(1)\big| = 1 \,, \hspace{10mm} (i=1,2,3,4)
\end{equation}
and are required to satisfy $\sgn F_1 = \sgn F_2$ and $\sgn F_3^\prime = \sgn F_4^\prime$ (which is equivalent to $f_i^2 \geq 0$).

The advantage of using $F_i$'s  can be already appreciated from the  \emph{BPS condition}
\begin{subequations}
\label{relationBPS}
\begin{eqnarray}
\label{relationBPSfi}
f_3 f_4 &=& H (f_1f_2)^{\prime} \,,
\end{eqnarray}
which attains in the language of $F_i$ much more elegant form
\begin{eqnarray}
\label{relationBPSFi}
F_3 &=& F_1 \,.
\end{eqnarray}
\end{subequations}
This condition will be assumed from now on.\footnote{The BPS condition \eqref{relationBPS} is in fact a generalization of the analogous condition presented already in \cite{Casana:2012un}. There, since $f_3^2=f_1^2$ and $f_4^2=f_2^2$ (meaning there are no terms $(\threevector{\phi} \innerdot D^\mu\threevector{\phi})^2$ and $(\threevector{\phi} \innerdot \threevector{F}^{\mu\nu})^2$ in the Lagrangian), it simplifies to $f_1^2 f_2^2 = H^2$.} 
%The reason is that, as shown in \cite{Benes:2023nsr}, under this condition we get, in the limit of vanishing potential, a BPS theory.

The \emph{raison d'\^{e}tre} of this condition is the following. Let us consider the static configuration of fields ($\partial^0 = 0$) and fix the gauge as $\threevector{A}^0 = 0$. Let us also define $\threevector{B}^i \equiv -\frac{1}{2} \epsilon^{ijk} \threevector{F}^{jk}$. Then, if the BPS condition \eqref{relationBPS} holds and the first-order BPS equations of motion
\begin{subequations}
\label{BPSeqnFiseparate}
\begin{eqnarray}
D^i\threevector{n}
&=& \frac{1}{vg} \frac{1}{F_2} \Big[\threevector{B}^i
- (\threevector{n}\innerdot\threevector{B}^i) \threevector{n}\Big]
\,,
%\\
%\frac{\partial^i H}{H}
%&=& \frac{1}{vg} \frac{1}{HF_4^\prime} (\threevector{n}\innerdot\threevector{B}^i)
%\,,
\\
\partial^i H
&=& \frac{1}{vg} \frac{1}{F_4^\prime} (\threevector{n}\innerdot\threevector{B}^i)
\,,
\end{eqnarray}
\end{subequations}
or equivalently and more compactly
\begin{eqnarray}
\label{BPSeqnFi}
D^i \threevector{\phi}
&=&
\frac{H}{g} \bigg[
\frac{1}{F_2} \bigg(\threevector{B}^i
- \frac{\threevector{\phi} \innerdot \threevector{B}^i}{\threevector{\phi}^2}\threevector{\phi}\bigg)
+ \frac{1}{H F_4^\prime} \frac{\threevector{\phi} \innerdot \threevector{B}^i}{\threevector{\phi}^2} \threevector{\phi}
\bigg]
\,,
\hspace{10mm}
\end{eqnarray}
are satisfied, the energy density is given solely by the total derivative term (provided $V \to 0$), as required:
\begin{eqnarray}
\label{EdensTotDer}
\mathcal{E} &=&
\partial^i \bigg(\frac{F_1}{gH} \threevector{\phi} \innerdot \threevector{B}^i\bigg)
\,.
\end{eqnarray}

\subsection{Spherical symmetry}
\label{sphericalsymm}

Let us now specialize to spherically symmetric field configurations. We consider the standard \qm{hedgehog} Ansatz
\begin{equation}
\label{ansatz}
\phi_a \ = \ v H \frac{x_a}{r} \,,
\hspace{8mm}
A^i_a \ = \ -\frac{1}{r^2} \varepsilon_{abi} x_b (1-K) \,,
\hspace{8mm}
\end{equation}
where the form factors $H$ and $K$ are functions of $r = |\threevector{r}|$ and satisfy the  boundary conditions
\begin{equation}
\label{boundarycondHK}
H(\infty) \ = \ 1 \,,
\hspace{10mm}
K(\infty) \ = \ 0 \,,
\end{equation}
that are necessary for the convergence of the energy. Notice that we have deliberately chosen $H(\infty) = + 1$ instead of $-1$; this, together with the choice
\begin{equation}
\label{f1f2sign}
F_1(1) \ = \ F_2(1) \ = \ +1 \,,
\end{equation}
implies that we are considering, without loss of generality, only monopoles and not anti-monopoles, as argued in~\cite{Benes:2023nsr}.

Let us stress that, in contrast to other authors, we do not impose the conditions $H(0) = 0$ and $K(0) = 1$ to ensure regularity of $\threevector{\phi}$ and $\threevector{A}^i$ in the origin. As we discussed in detail already in \cite{Benes:2023nsr}, we adopt the viewpoint that a singularity in a \emph{field} is not necessarily a problem. Due to the reparameterization invariance of a field theory, we view such a singularity merely as a kind of \emph{coordinate singularity} that can be cured by switching to better coordinates. In particular, a singularity in $\threevector{\phi}$ can be removed using the transformation \eqref{transformationhi} with, e.g., $\HtoH(\tilde H) = 1/\tilde H$. The case of $\threevector{A}^i$, while also in principle straightforward, is technically more complicated and we devote Sec.~\ref{SingularA} to it. What will worry us, however, will be singularities in \emph{physical} quantities like, e.g., the energy density.

Under the spherically symmetric Ansatz \eqref{ansatz} the BPS equations \eqref{BPSeqnFiseparate} turn into a system of two ordinary differential equations for $K$ and $H$:
\begin{subequations}
\label{logKH}
\begin{eqnarray}
\label{logK}
\partial_\rho(\log K) &=& - F_2(H)
\,,
%\\
%\label{logH}
%\partial_\rho H &=& 
%\frac{1-K^2}{\rho^2}
%\frac{1}{F_4^\prime(H)}
%\,,
\\
\label{logH}
\partial_\rho (\log H) &=& 
\frac{1-K^2}{\rho^2}
\frac{1}{H F_4^\prime(H)}
\,,
\end{eqnarray}
\end{subequations}
where we introduced a dimensionless radius
\begin{eqnarray}
\rho &\equiv& vgr \,.
\end{eqnarray}

Let us now turn our attention to the energy density \eqref{EdensTotDer}. We require that $\mathcal{E}$ is regular as $r \to 0$. The first reason for this is that we simply do not consider divergent $\mathcal{E}$ physical. Admittedly, this might seem a bit prejudicial, as a singularity of the type $1/r^2$ would still be integrable and would not jeopardize the finiteness of the total energy.

Nevertheless, there is a second, more formal reason. Since $\mathcal{E}$ is, in the BPS limit, given by a total derivative, we want the total energy (i.e., the monopole mass) $M = \int_{\mathbb{R}^3} \d^3 x \, \mathcal{E}$ to be given only by the surface term. (Which is proportional to a topological invariant -- the degree of the mapping of the manifold of spatial infinities to the vacuum manifold.) In other words, we want to deal with \emph{topological} monopoles. However, to achieve that (that is, to be allowed to apply the Gauss--Ostrogradsky divergence theorem), the vector field $V^i$ (and consequently, also $\mathcal{E}$ itself) in $\mathcal{E} = \partial^i V^i$ must be regular everywhere.

Under the spherically symmetric Ansatz \eqref{ansatz} the energy density \eqref{EdensTotDer} reads 
\begin{eqnarray}
\mathcal{E} &=&
\partial^i \bigg( \frac{v}{g} \frac{F_1 (1-K^2)}{r^3} x^i \bigg)
\,.
\end{eqnarray}
We see that, without making any assumptions about behavior of $K(r)$, the function $F_1(H(r))$ must vanish sufficiently fast as $r \to 0$,
\begin{eqnarray}
\label{F1H0is0}
F_1(H(0)) &=& 0 \,,
\end{eqnarray}
in order for $\mathcal{E}$ to be regular, as required.

Switching from $r$ to $\rho$ and factorizing out some constants, the energy density can be further simplified into
\begin{eqnarray}
\label{EdensitySphericalTot}
\frac{\mathcal{E}}{v^4g^2} &=&
\frac{\partial_\rho \big[F_1 (1-K^2) \big]}{\rho^2}
\\ &=&
\label{EdensitySphericalFi}
2 F_1 F_2 \frac{K^2}{\rho^2}
+
\frac{F_1^\prime}{F_4^\prime}
\frac{(1-K^2)^2}{\rho^4}
\,,
\end{eqnarray}
where, on the second line, we used the BPS equations to eliminate the derivatives with respect to $\rho$. Using \eqref{EdensitySphericalTot} and \eqref{F1H0is0}, the mass of the monopole
\begin{eqnarray}
M &=&
\int_{\mathbb{R}^3}\!\d^3 x \, \mathcal{E}
\ = \ 
\frac{4 \pi v}{g} \int_{0}^{\infty}\!\d\rho \, \rho^2 \frac{\mathcal{E}}{v^4 g^2}
\,,
\end{eqnarray}
follows immediately as
\begin{eqnarray}
M &=&
\frac{4 \pi v}{g} F_1\big(H(\infty)\big) \, \big(1-K^2(\infty)\big)
\,.
\end{eqnarray}
Finally, using \eqref{boundarycondHK} and \eqref{f1f2sign}, we get the neat result
\begin{eqnarray}
M &=& \frac{4 \pi v}{g} \,.
\end{eqnarray}

\section{Solving the BPS equations}
\label{InvertibleCaseGeneral}

\subsection{The solution}

There are two general classes of models for which the BPS equations \eqref{logKH} can be solved analytically. The first class (which leads to a generalization of the usual magnetic monopole of the 't~Hooft--Polyakov type) was studied in detail in \cite{Benes:2023nsr}. Recall that, in this case, explicit solutions were found under the crucial assumption that $F_2$ is an \emph{invertible} function (and with the additional constraint $F_2 = F_4$).

However, there is also another class of models. Let us consider the situation when $F_2$ is \emph{not invertible}. In particular, let us focus, for simplicity, on the \qm{least} invertible function $F_2$, that is, on a constant $F_2$. Due to the normalization \eqref{f1f2sign} there is only one possibility:
\begin{eqnarray}
\label{F2is1}
F_2 &=& 1 \,.
\end{eqnarray}
Together with the BPS condition \eqref{relationBPS} and $V=0$, this means that we are considering the Lagrangian
\begin{eqnarray}
\eL &=&
\frac{v^2}{2} \bigg[
F_1 \bigg(\frac{(D^\mu \threevector{\phi})^2}{\threevector{\phi}^2}-\frac{(\threevector{\phi} \innerdot D^\mu \threevector{\phi})^2}{\threevector{\phi}^4}\bigg)
+ H^2 F_1^\prime F_4^\prime \, \frac{(\threevector{\phi} \innerdot D^\mu \threevector{\phi})^2}{\threevector{\phi}^4}
\bigg]
\nonumber \\ && {}
- \frac{1}{4g^2} \bigg[
F_1 \bigg((\threevector{F}^{\mu\nu})^2-\frac{(\threevector{\phi} \innerdot \threevector{F}^{\mu\nu} )^2}{\threevector{\phi}^2}\bigg)
+ \frac{F_1^\prime}{F_4^\prime} \frac{(\threevector{\phi} \innerdot \threevector{F}^{\mu\nu})^2}{\threevector{\phi}^2}
\bigg]
\,.
\nonumber \\ && {}
%- V(\threevector{\phi}^2)
\end{eqnarray}

The solution to \eqref{logK} immediately follows as
\begin{eqnarray}
\label{solutionK}
K(\rho) &=& \xi\exp(-\rho) \,,
\end{eqnarray}
where $\xi \in \mathbb{R}$ is a constant of integration. The second BPS equation, \eqref{logH}, can be now solved as
\begin{eqnarray}
\label{H2intermsofF}
%H &=& F_4^{-1}(1-H_0) \,,
%\\
H(\rho) &=& F_4^{-1}\big(F_4(1)-\lambda(\rho)\big) \,,
\end{eqnarray}
where $F_4(1) = \pm1$ (due to \eqref{Finormalization}) is a constant of integration and where we introduced the auxiliary function
\begin{widetext}
%\begin{subequations}
\begin{eqnarray}
\label{Ei_H0}
\lambda(\rho) &\equiv& 
\int^{\infty}_{\rho} \!\d \rho \frac{1-K^2}{\rho^{2}}
%\\ &=&
%\ = \ 
=
\frac{1-\xi^2 \e^{-2\rho}}{\rho} - 2 \xi^2 \Ei(-2\rho)
%\\ &=&
%\ = \ 
=
\begin{cases}
\displaystyle
\frac{1-\xi^2}{\rho}
%& \\
%\displaystyle 
%\hspace{4mm}
%- 2\xi^2
%\bigg[
%\log(2\e^{\gamma_{\mathrm{E}}-1}\rho)
%+ \sum_{n=1}^\infty \frac{(-2\rho)^n}{n(n+1)n!}
%\bigg]
- 2\xi^2
\bigg[
\log(2\rho)
+ \EM-1
+ \sum_{n=1}^\infty \frac{(-2\rho)^n}{n \, (n+1)!}
\bigg]
\,,
& (\rho \to 0)
\\[1em]
\displaystyle
\frac{1}{\rho} \bigg[ 1 + \xi^2 \e^{-2\rho}\sum_{n=1}^{\infty} \frac{n!}{(-2\rho)^n} \bigg]
\,,
& (\rho \to \infty)
\end{cases}
\nonumber \\ &&
\end{eqnarray}
%\end{subequations}
\end{widetext}
where
%\begin{subequations}
\begin{eqnarray}
\Ei(x)
&\equiv&
%- \int_{-x}^{\infty} \frac{\e^{-t}}{t} \, \d t
\int^{x}_{-\infty}\! \frac{\e^{t}}{t} \, \d t
%\\ &=&
%\ = \ 
=
\begin{cases}
\displaystyle 
\log|x| + \EM + \sum_{n=1}^\infty \frac{x^n}{n\,n!}
\,,
& (x \to 0)
\\[1em]
\displaystyle 
\frac{\e^{x}}{x} \sum_{n=1}^{\infty} \frac{n!}{x^n}
\,,
& (x \to -\infty)
\end{cases}
%\nonumber \\ &&
\end{eqnarray}
%\end{subequations}
is the \emph{exponential integral}\footnote{
Other notations for the exponential integral, commonly used in the literature, are
%$\Ei(x) = -\mathrm{E}_1(-x) = -\mathrm{Ei}(1,-x)$.
$\mathrm{E}_1(x) = \mathrm{Ei}(1,x) = -\Ei(-x)$.
} and $\EM = 0.57721\ldots$ is the Euler--Mascheroni constant.
% Notice that the boundary conditions $K(\infty) = 0$ and $H(\infty) = 1$ are satisfied automatically, for \emph{any value} of $\xi$.
%We plot $\lambda(\rho)$ in Fig.~\ref{fig_Ei_HO}.

%Would it be possible that the two lines of the equation (34) be numbered respectively as (34a) and (34b)? I.e., like when using the \begin{subequations}...\end{subequations} environment? I didn't manage to achieve it myself.

\begin{figure}[t]
\begin{center}
\includegraphics[width=\linewidth]{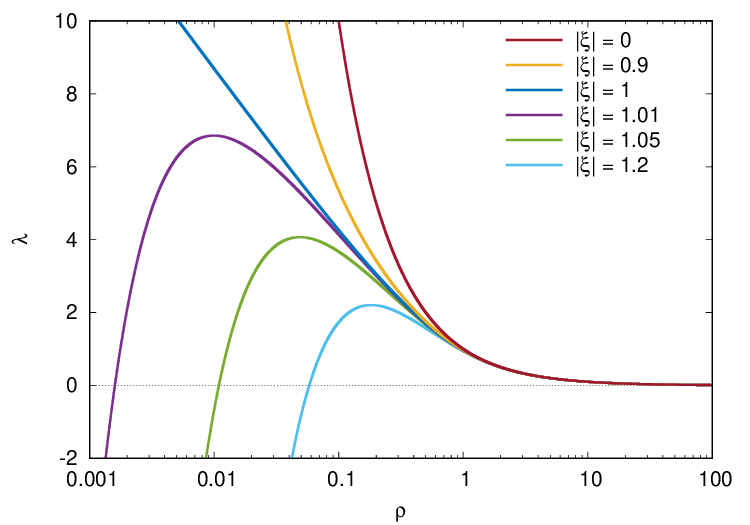}
\caption{The function $\lambda(\rho)$, Eq.~\eqref{Ei_H0}, for various values of the parameter $|\xi|$. Notice the different behavior for $|\xi| \leq 1$ and for $|\xi| > 1$.}
\label{fig_Ei_HO}
\end{center}
\end{figure}

As illustrated in Fig.~\ref{fig_Ei_HO}, the function $\lambda(\rho)$ always vanishes for $\rho \to \infty$. The behavior of $\lambda(\rho)$ near origin, however, depends on $\xi$:
\begin{itemize}
\item If $\xi^2 \leq 1$, then $\lambda(\rho \to 0) = +\infty$ and for $\rho \in (0,\infty)$ it is monotonically decreasing.
\item If $\xi^2 > 1$, then $\lambda(\rho \to 0) = -\infty$ and there is a maximum at $\rho_{\mathrm{max}} = \log |\xi|$ with the value $\lambda(\rho_{\mathrm{max}}) = -2 \xi^2 \Ei(-\log \xi^2)$.
\end{itemize}

\subsection{The parameter $\xi$}

The presence of the constant of integration $\xi$ is suspicious: One would expect $\xi$ to be fixed to some particular value by, e.g., the boundary conditions or other requirements. So let's see if this is possible.

To constrain the possible values of $\xi$, we turn to the energy density $\mathcal{E}$ and require that it be regular and positive everywhere and that the integral $\int_0^\infty \d\rho \, \rho^2 \frac{\mathcal{E}}{v^4g^2}$ be finite. 

To that end, let us first write the energy density, corresponding to the solution \eqref{solutionK}, \eqref{H2intermsofF}, in the form
\begin{eqnarray}
\label{EdensityNonInv}
\frac{\mathcal{E}}{v^4g^2}
&=&
2 F(\lambda) \frac{\xi^2\e^{-2\rho}}{\rho^2}
- \frac{\d F(\lambda)}{\d \lambda} \frac{(1-\xi^2\e^{-2\rho})^2}{\rho^4}
%\\ &=&
%2 F(\lambda) \frac{\xi^2}{\rho^2} \big(1-2\rho+\frac{1}{2}(-2\rho)^2+\mathcal{O}(\rho^3)\big)
%\nonumber \\ && {}
%- \frac{\d F(\lambda)}{\d \lambda} \frac{(1-\xi^2\e^{-2\rho})^2}{\rho^4}
\,,
\end{eqnarray}
where $\lambda = \lambda(\rho)$ is given by \eqref{Ei_H0} and where we defined, for convenience, the function
\begin{eqnarray}
\label{Ffirst}
%F(H_0) &\equiv& F_1\big(F_4^{-1}(1-H_0)\big) \,.
%\\
F(\lambda) &\equiv& F_1\Big(F_4^{-1}\big(F_4(1)-\lambda\big)\Big) \,.
\end{eqnarray}
Notice that it is obviously normalized as
\begin{eqnarray}
F(0) &=& 1
\end{eqnarray}
%$F(0) = 1$, 
and, perhaps less obviously, it satisfies
\begin{equation}
F(\lambda) \ \geq \ 0
\hspace{6mm}
\mbox{and}
\hspace{6mm}
F^{\prime}(\lambda) \ \leq \ 0
\,,
\hspace{6mm}
\mbox{(for all $\lambda$)}
\end{equation}
%$F \geq 0$ and $\d F / \d \lambda \leq 0$ (for all $\lambda$), 
so that both terms in \eqref{EdensityNonInv} are separately non-negative.

By the way, the fact that the energy density can be written in terms of a \emph{single} function is not a coincidence. In fact, the function $F(\lambda)$, with $\lambda$ understood now as a function of $H$ (by inverting \eqref{H2intermsofF}), fully and uniquely specifies a given model (i.e., the physics), up to the redundancy mentioned in Sec.~\ref{forminvariance}. We will elaborate a bit more on this point in Sec.~\ref{redundancyagain}.

%\begin{equation}
%F \ \geq \ 0 \,,
%\hspace{10mm}
%\frac{\d F}{\d \lambda} \ \leq \ 0
%\end{equation}

For large $\rho$ the energy density goes like
\begin{eqnarray}
\frac{\mathcal{E}}{v^4g^2}
&=&
- \frac{\d F(\lambda)}{\d \lambda}\bigg|_{\lambda = 0} \, \frac{1}{\rho^4}
+ \ldots
\end{eqnarray}
It falls off just fast enough for the integral $\int_0^\infty \d\rho \, \rho^2 \frac{\mathcal{E}}{v^4g^2}$ to be convergent at the upper limit, regardless of the value of $\xi$. This is consistent with the fact that the boundary conditions \eqref{boundarycondHK}
(that are, after all, designed exactly to ensure the convergence of the energy density at the upper limit)
are satisfied by the solution \eqref{solutionK} and \eqref{H2intermsofF} automatically, for any value of $\xi$.
Thus, the regularity of the energy density at large distances doesn't give any constraint on the possible values of $\xi$.

The only other region where the energy density \eqref{EdensityNonInv} might be singular is at the origin. To investigate its behavior at $\rho \to 0$, let us first assume that $\xi^2 > 1$. The behavior of $\lambda(\rho)$ is then such that $\lambda(\rho \to 0) \to -\infty$, see Fig.~\ref{fig_Ei_HO}. At this point, there are three requirements for the function $F(\lambda)$. First, $F(\lambda \to -\infty) \to 0$ so that the first term is regular. Second, $F > 0$ (for all $\lambda$) so that the first term is positive. Third, $\d F / \d \lambda < 0$ (for all $\lambda$) so that the second term is positive. However, this is a contradiction, as these three requirements cannot be met simultaneously. Therefore it follows that the case $\xi^2 > 1$ is not possible and we are left with
\begin{eqnarray}
\xi^2 &\leq& 1 \,,
\end{eqnarray}
which doesn't lead to any contradiction and, as such, is in principle allowed.

The condition $\xi^2 \leq 1$ is the furthest we can go in constraining possible values of $\xi$ in a model-independent way. Accordingly, it is only necessary, but by no means sufficient condition. To proceed, one has to specify a particular model, i.e., choose the function $F(\lambda)$ which must satisfy
\begin{equation}
F(\infty) \ = \ 0
\hspace{6mm}
\mbox{and}
\hspace{6mm}
F^{\prime}(\infty) \ = \ 0
\end{equation}
to have energy density regular at the origin (recall that $\lambda \to \infty$ corresponds to $\rho \to 0$).
This requirement can further constrain $\xi$. In this regard, typically one of the three possibilities can happen:
\begin{itemize}
\item All values $\xi^2 \leq 1$ are allowed.
\item Only the values $\xi^2 < 1$ are allowed.
\item No value of $\xi$ is allowed.
\end{itemize}
Needless to say, the last possibility means that there is no finite-energy monopole solution for that model.

To illustrate this, let us consider the case when $F(\lambda)$ behaves for large $\lambda$ (i.e., small $\rho$) like a power function:
\begin{eqnarray}
F(\lambda\to\infty) &=& A \lambda^{-N} + \ldots\,,
\end{eqnarray}
where $N > 0$. The leading divergent term in energy density is $\mathcal{E} / v^4g^2 = NA \lambda^{-N-1} (1-\xi^2\e^{-2\rho})^2 / \rho^4$. Let us first assume $\xi^2 = 1$. In this case, $\lambda$ behaves like a $\log\rho$. However, any negative power of $\log\rho$ cannot compete with $1/\rho^2$, therefore $\xi^2 = 1$ is not viable, as it would lead to divergent $\mathcal{E}$. So let's assume $\xi^2 < 1$. Now the energy density goes near the origin like $\mathcal{E} / v^4g^2 = NA (1-\xi^2)^{1-N} \rho^{N-3}$. Obviously, this is regular as long as $N \geq 3$ for any $\xi^2 < 1$. To summarize:
\begin{itemize}
\item $N < 3$: No $\xi$ is allowed.
\item $N \geq 3$: $\xi^2 < 1$ are allowed.
\end{itemize}

Other and more complete examples will be presented in Sec.~\ref{examples}. Nevertheless, the basic conclusion is always the same: Whenever there exists, for a given model, a physical monopole solution (i.e., with regular energy density and with finite total energy, the mass), there exists a whole continuous family of such solutions labeled by $\xi$ taking values from a unit interval.

The important point is that $\xi$ is a physical, measurable parameter since the shape of the energy density explicitly depends on it. Within the same model, the different $\xi$ correspond to different energy density profiles. At the same time, monopoles with different $\xi$ have the same mass $M = 4\pi v/g$.\footnote{
This is a consequence of our requirement (expressed technically by $F(\infty) = 0$) that $\mathcal{E}$ be regular in the origin and, accordingly, that the monopole in question is \emph{topological}. If, on the contrary, the energy density had a singularity of the type $1/\rho^2$ (that is, if $F(\infty) \neq 0$), the total energy would be $M = \frac{4 \pi v}{g}[ 1 + F(\infty) (\xi^2-1)]$. In other words, the mass of the corresponding \emph{non-topological} monopole would explicitly depend on $\xi$.
} In this sense the parameter $\xi$ can be interpreted as an internal degree of freedom of the monopole, or as a moduli space parameter.

\subsection{Exploiting the redundancy}
\label{redundancyagain}

Our Lagrangian (or more precisely, the class of Lagrangians) \eqref{lagrangian} was parameterized by four form-functions $f_i^2$, or equivalently, by $F_i$. (We are leaving aside the nullified potential.) Out of these four functions, one was eliminated by the BPS condition $F_3 = F_1$ and another one by the additional condition $F_2 = 1$, so we got left with only two independent functions, $F_1$ and $F_4$, that parameterize our class of Lagrangians.

Recall, however, the redundancy of this parameterization, discussed in Sec.~\ref{forminvariance}: Due to the form-invariance of the Lagrangian under rescaling $H \to \HtoH(H)$, one of the form-functions is redundant, and it can be set to (almost) any prescribed function without affecting the physics. It follows that of the two remaining functions, only one of them (or just a single combination of them), let us call it $F$, uniquely describes the physics. The other one, let us call it $G$, merely parameterizes the redundancy of our description, or in other words, serves as a \qm{coordinate} on the space of physically equivalent Lagrangians and as such is non-physical.

For instance, we can identify
\begin{subequations}
\label{FG}
\begin{eqnarray}
F_4(H) &=& G(H) \,,
\\
\label{Fnew}
F_1(H) &=& F\big(\Lambda(H)\big) \,,
\end{eqnarray}
\end{subequations}
where
\begin{eqnarray}
\Lambda(H) &\equiv& G(1)-G(H) \,.
\end{eqnarray}
The solution is now $K(\rho) = \xi\e^{-\rho}$ and
\begin{eqnarray}
\label{HrhoG}
H(\rho) &=& G^{-1}\big( G(1)-\lambda(\rho) \big) \,.
\end{eqnarray}
Note that the definition \eqref{Fnew} of $F$ is equivalent to the former definition \eqref{Ffirst} of $F$, as for the solution \eqref{HrhoG} it is $\Lambda(H(\rho)) = \lambda(\rho)$. In the language of $F$ and $G$ of \eqref{FG} the class of Lagrangians considered in this text reads
\begin{eqnarray}
\label{lagFG}
\hspace{-20mm}
\eL &=&
\frac{v^2}{2} \bigg[
F \bigg(\frac{(D^\mu \threevector{\phi})^2}{\threevector{\phi}^2}
- \frac{(\threevector{\phi} \innerdot D^\mu \threevector{\phi})^2}{\threevector{\phi}^4}\bigg)
\nonumber
-
F^\prime \big(H G^\prime\big)^2 \frac{(\threevector{\phi} \innerdot D^\mu \threevector{\phi})^2}{\threevector{\phi}^4}
\bigg]
\nonumber \\ && {}
- \frac{1}{4g^2} \bigg[
F \bigg((\threevector{F}^{\mu\nu})^2
- \frac{(\threevector{\phi} \innerdot \threevector{F}^{\mu\nu})^2}{\threevector{\phi}^2}\bigg)
- F^\prime \frac{(\threevector{\phi} \innerdot \threevector{F}^{\mu\nu})^2}{\threevector{\phi}^2}
\bigg]
\,,
\nonumber \\ &&
\end{eqnarray}
where $F \equiv F(\Lambda(H))$ and $F^\prime = \d F / \d \Lambda$. In this form, all Lagrangians with the same $F$ are physically equivalent, regardless of possibly different $G$.

\subsection{The behavior of the gauge potential}
\label{SingularA}

With our solution $K = \xi \e^{-\rho}$ the gauge potential $\threevector{A}^i$ goes near the origin like
\begin{eqnarray}
A^i_a &=&
- \varepsilon_{abi} x_b \frac{1-K}{r^2}
\ = \ 
- \varepsilon_{abi} x_b \bigg[ \frac{1-\xi}{r^2} + \frac{\xi vg}{r} + \mathcal{O}(1)\bigg] \,.
\nonumber \\ &&
\end{eqnarray}
We see that $\threevector{A}^i$ is singular in the origin for \emph{any value} of $\xi$. If $\xi = 1$, the singularity is somewhat lessened, but still present, as the value of $\threevector{A}^i$ at the origin depends on the direction from which the limit is taken.\footnote{
The case of $\xi = 1$ has already been studied in \cite{Bazeia:2018fhg}, but without noticing the issue of singularity in $\threevector{A}^i$. Cf.~also the remark after equations \eqref{LagPowSimpSolution}.
} Thus, the solution could be, at first sight, deemed unphysical.

On general grounds, and in the spirit of the philosophy adopted in \cite{Benes:2023nsr}, it can be argued that this is not a problem, since the gauge fields themselves are non-physical. What matters are singularities in measurable quantities, e.g., the energy density, whose regularity has been carefully discussed and maintained in the previous sections.

Nevertheless, let us show more explicitly that the singularity in the gauge field is harmless. In fact, it can be transformed away using the transformation of the gauge field $\threevector{A}_\mu \to \tilde{\threevector{A}}_\mu$ as\footnote{This is a special case of the more general transformation considered in \cite{Benes:2023nsr} (Appendix~A therein) with $h(\tilde H) = k(\tilde H) = 0$ and $\alpha(\tilde H) = \tilde H$.}
\begin{eqnarray}
\label{transformationofA}
\threevector{A}_\mu
&=&
\tilde{\threevector{A}}_\mu
+
\big(\ell-1\big) \frac{\threevector{\phi} \times \tilde{D}_\mu \threevector{\phi}}{\threevector{\phi}^2}
\,,
\end{eqnarray}
where $\tilde{D}_\mu \threevector{\phi} \equiv \partial_\mu \threevector{\phi} + \tilde{\threevector{A}}_\mu \times \threevector{\phi}$ and $\ell = \ell(H)$ is some function. The point is that, as we show in \cite{Benes:2023nsr}, this transformation preserves the spherically symmetric form \eqref{ansatz} of the gauge field and only rescales the form factor $K$ as
\begin{equation}
K \ \to \ 
\tilde{K}
\ \equiv \ 
\frac{K}{\ell}
\,.
\end{equation}
However, since \eqref{transformationofA} is not a gauge transformation, but rather a field redefinition, the form of the Lagrangian is not preserved; instead, new interaction terms involving up to four derivatives emerge. Nevertheless, now we may demand $\tilde{K} = 1 + \mathcal{O}(r^2)$ in order for $(1-\tilde{K})/r^2$ to stay regular as $r \to 0$. It follows that $\ell$ must behave like $\ell(H(\rho)) = \xi [1-\rho+\mathcal{O}(\rho^2) ]$. Next, since we want $\ell$ expressed as a function of $H$ (rather than of $\rho$), we have to invert $H(\rho)$ for small $\rho$ and plug it back into the previous form of $\ell(H(\rho))$. This step now depends on the precise form of $H(\rho)$. For instance, if it is possible to expand $H(\rho) = H(0) + \rho H^{\prime}(0) + \mathcal{O}(\rho^2)$ with $H^{\prime}(0) \neq 0$, then any $\ell$ of the form
\begin{eqnarray}
\ell(H) &=&
\xi \bigg[
1
- \frac{H-H(0)}{H^{\prime}(0)}
+ \mathcal{O}\Big(\big(H-H(0)\big)^2\Big)
\bigg]
\end{eqnarray}
will do the desired job of removing the singularity from the gauge field.

\section{Examples}
\label{examples}

\subsection{Power-function Lagrangian -- Special case}
\label{subsec:power}

\begin{figure*}[t]
\centering
\begin{minipage}{\textwidth}
\centering
\includegraphics[width=0.5\textwidth]{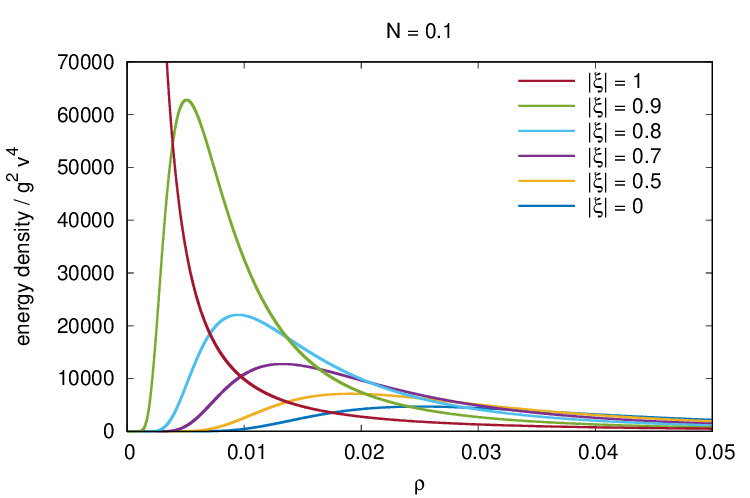}%
\includegraphics[width=0.5\textwidth]{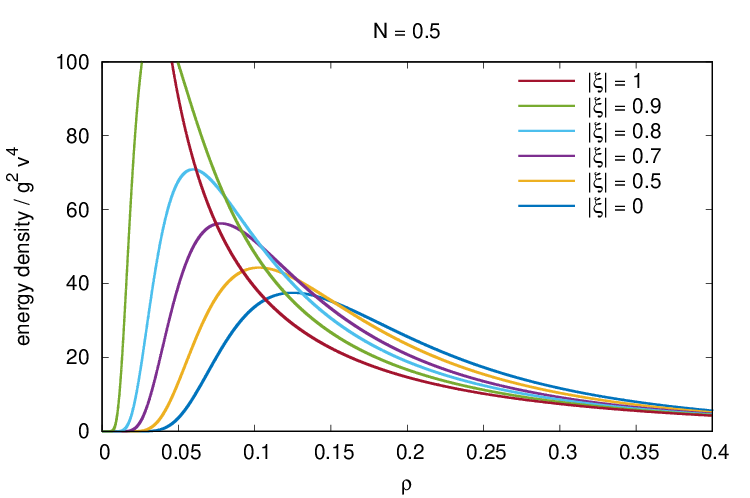}
\end{minipage}
\begin{minipage}{\textwidth}
\centering
\includegraphics[width=0.5\textwidth]{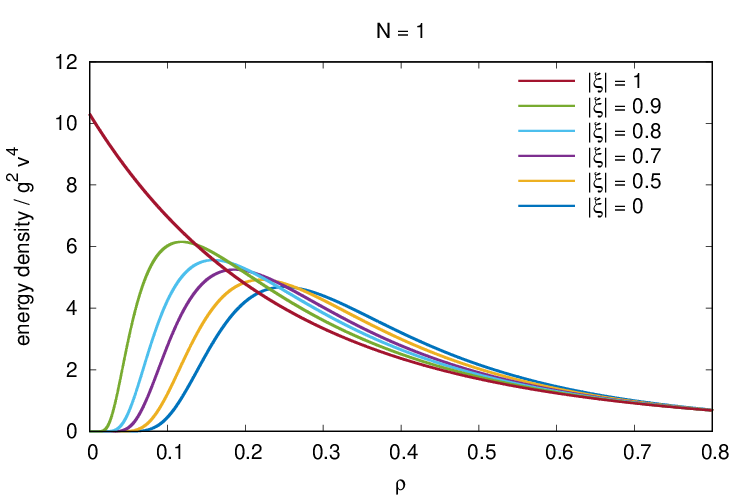}%
\includegraphics[width=0.5\textwidth]{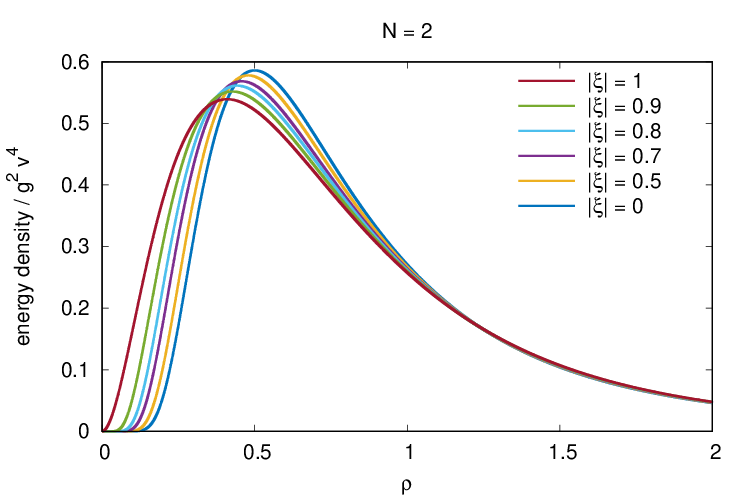}
\end{minipage}
\caption{Energy densities \eqref{edensitypownoninv} for a single monopole solution of the power-function theory \eqref{LagPowSimp} for $N = n/m = 0.1,\, 0.5,\, 1,\, 2$ and for various values of $|\xi| \leq 1$. In the case of $N < 1$ (top panels) the limit value $|\xi| = 1$ is included just to show that the energy density indeed diverges in this case. Notice that dependence of the energy density profile on $\xi$ is stronger for smaller $N$.}
\label{plot_noninv_pow}
\end{figure*}

As the simplest example, let's consider the Lagrangian
\begin{eqnarray}
\label{LagPowSimp}
\eL &=&
\frac{v^2}{2} H^n \bigg[
\frac{(D^\mu \threevector{\phi})^2}{\threevector{\phi}^2}
+ (nm-1) \frac{(\threevector{\phi} \innerdot D^\mu \threevector{\phi})^2}{\threevector{\phi}^4}
\bigg]
\nonumber \\ && {}
- \frac{1}{4g^2} H^n \bigg[
(\threevector{F}^{\mu\nu})^2
+ \bigg(\frac{n}{m}-1\bigg) \frac{(\threevector{\phi} \innerdot \threevector{F}^{\mu\nu} )^2}{\threevector{\phi}^2}
\bigg]
\,,
\hspace{8mm}
\end{eqnarray}
where $m$, $n$ are some constants. In the language of $F_i$ this corresponds to
\begin{equation}
F_1 = F_3 = H^{n} \,,
\hspace{5mm}
F_2 = 1 \,,
\hspace{5mm}
F_4 = 1 + m \log H \,.
\end{equation}
Both conditions \eqref{relationBPS} and \eqref{F2is1} are therefore satisfied, so we can call up the methods developed above and find the monopole solution
\begin{subequations}
\label{LagPowSimpSolution}
\begin{eqnarray}
K &=& \xi \e^{-\rho} \,,
\\
H &=& 
\exp \Bigg[-\frac{1}{m} \bigg(
\frac{1-\xi^2 \e^{-2\rho}}{\rho} - 2 \xi^2 \Ei(-2\rho)
\bigg)\Bigg]
\,.
\hspace{10mm}
\end{eqnarray}
\end{subequations}

Let us briefly discuss the special case $m = n = \pm 1$, when the \qm{new} kinetic terms \eqref{termsKinNew}, $(\threevector{\phi} \innerdot D^\mu \threevector{\phi})^2$ and $(\threevector{\phi} \innerdot \threevector{F}^{\mu\nu})^2$, are absent from the Lagrangian \eqref{LagPowSimp}. In fact, such a Lagrangian is actually the only Lagrangian lacking these terms and at the same time belonging to our class of models possessing monopole solutions with the parameter $\xi$. Interestingly, this particular model (with $m = n = +1$) has been already studied in \cite{Bazeia:2018fhg}, with the resulting solution (dubbed a \qm{small monopole}) being in agreement with our solution \eqref{LagPowSimpSolution}. However, the authors of \cite{Bazeia:2018fhg} considered, in our language, only the $\xi = 1$ case.

Next, we have, using \eqref{Ffirst},
\begin{eqnarray}
F(\lambda) &=& \e^{-N\lambda} \,,
\end{eqnarray}
where we denoted
\begin{eqnarray}
N &\equiv& \frac{n}{m} \,,
\end{eqnarray}
so the energy density follows as
\begin{eqnarray}
\label{edensitypownoninv}
\frac{\mathcal{E}}{v^4g^2}
&=&
\e^{-N\lambda} \bigg[2 \frac{\xi^2\e^{-2\rho}}{\rho^2}
+ N \frac{(1-\xi^2\e^{-2\rho})^2}{\rho^4}\bigg]
\,.
\hspace{5mm}
\end{eqnarray}
The fact that the energy density depends on the parameters $m$, $n$ only through their particular combination $n/m$, is, of course, a manifestation of the mentioned redundancy of the Lagrangian: The Lagrangians may have different $m$, $n$, but as long as they have the same ratio $n/m$, they are physically equivalent, as exemplified in \eqref{edensitypownoninv}.

Needless to say, the class of Lagrangians \eqref{LagPowSimp} labeled by $m$, $n$ doesn't exhaust all physically equivalent Lagrangians (i.e., those that lead to $F(\lambda) = \e^{-N\lambda}$ with the same $N$); there are infinitely many more of them.

After all, instead of starting with the Lagrangian \eqref{LagPowSimp}, we could have turned our analysis \qm{upside down}, in the spirit of Sec.~\ref{redundancyagain}: First, to choose the function $F(\Lambda)$ (in our case, $F(\Lambda) = \exp(-N\Lambda)$) that defines the physics and discuss the monopole's properties like the energy density profile. And \emph{only then} to choose a function $G(H)$ (in our case, $G(H) = 1 + m \log H$) to be able to write down, via \eqref{lagFG}, a Lagrangian and, correspondingly, to obtain the particular solution $H(\rho)$.

Let us now investigate the possible values of $N$ and $\xi$. Since $\lambda(0) = \infty$, it must be in any case $N>0$ for the energy density to be regular in the origin. However, this is only a sufficient condition. To proceed, let's expand $\mathcal{E}$ for small $\rho$:
\begin{widetext}
\begin{eqnarray}
\frac{\mathcal{E}}{v^4g^2}
&=&
\begin{cases}
\displaystyle 
N (1-\xi^2)^2
\big(2 \e^{\EM-1}\big)^{2\xi^2N}
\rho^{2(\xi^2N-2)}
\e^{-N\frac{1-\xi^2}{\rho}}
\big[ 1 + \mathcal{O}(\rho) \big]
\,,
& (\xi^2 < 1)
\\[1em]
\displaystyle 
2 \big(2 \e^{\EM-1}\big)^{2N}
\big(1+2N\big)
\rho^{2(N-1)}
\big[ 1 + \mathcal{O}(\rho) \big]
\,.
& (\xi^2 = 1)
\end{cases}
\end{eqnarray}
\end{widetext}
There are therefore three regimes:
\begin{itemize}
\item Case $0 < N < 1$: The energy density in the origin is singular for $\xi^2 = 1$, but regular (and even vanishing) for any $\xi^2 < 1$.
\item Case $N = 1$: The energy density is regular for all $\xi^2 \leq 1$. For $\xi^2 < 1$ it is vanishing at the origin, but for $\xi^2 = 1$ it takes on the non-zero value $\mathcal{E}(0)/v^4g^2 = 24 \, \e^{2(\EM-1)} \approx 10.3$.
\item Case $N > 1$: The energy density is regular and vanishing in the origin for all $\xi^2 \leq 1$.
\end{itemize}
See Fig.~\ref{plot_noninv_pow}, where we plot energy densities for $N$ from all these three regimes ($N = 0.1,\, 0.5,\, 1,\, 2$).
To summarize, the allowed values of $\xi^2$ depend on $N$ as
\begin{subequations}
\begin{eqnarray}
0 < N < 1 &\hspace{5mm}\Rightarrow\hspace{5mm}& \xi^2    < 1 \,, \\
N \geq  1 &\hspace{5mm}\Rightarrow\hspace{5mm}& \xi^2 \leq 1 \,,
\end{eqnarray}
\end{subequations}
in order to have $\mathcal{E}$ finite.

Thus, indeed, we have found that a residual dependence on the constant of integration $\xi$, unconstrained by any constraint, is left in the solution: $\xi$ controls the shape of the energy density profile, or, loosely speaking, the \qm{radius} of the monopole. Consequently, $\xi$ is a physical (measurable) parameter. Moreover, $\xi$ can vary smoothly within a continuous interval, between $-1$ and $+1$.

\subsection{Power-function Lagrangian -- General case}

The previous Lagrangian \eqref{LagPowSimp} can be slightly generalized, by adding a new parameter $k$:
\begin{eqnarray}
\label{LagPowGen}
\eL &=&
\frac{v^2}{2} H^n \bigg[
\frac{(D^\mu \threevector{\phi})^2}{\threevector{\phi}^2}
+ \bigg(\frac{nm}{H^{k}}-1\bigg) \frac{(\threevector{\phi} \innerdot D^\mu \threevector{\phi})^2}{\threevector{\phi}^4}
\bigg]
\nonumber \\ && {}
- \frac{1}{4g^2} H^n \bigg[
(\threevector{F}^{\mu\nu})^2
+ \bigg(\frac{n}{m}H^k-1\bigg) \frac{(\threevector{\phi} \innerdot \threevector{F}^{\mu\nu} )^2}{\threevector{\phi}^2}
\bigg]
\,.
\hspace{8mm}
\end{eqnarray}
The Lagrangian \eqref{LagPowSimp} corresponds to $k = 0$. In fact, with arbitrary $k$, we now have the most general Lagrangian with all form-functions $f_i^2$ being power functions that satisfy all our requirements. The reason why we are considering the case $k \neq 0$ separately is that the corresponding energy density has slightly different properties than in the $k = 0$ case.

The Lagrangian \eqref{LagPowGen} corresponds to
\begin{equation}
F_1 = F_3 =  H^{n} \,,
\hspace{5mm}
F_2 = 1 \,,
\hspace{5mm}
F_4 = 1 + \frac{m}{k} \bigg(1-\frac{1}{H^{k}}\bigg)
\,,
\end{equation}
the conditions \eqref{relationBPS} and \eqref{F2is1} are again satisfied and the monopole solution readily follows as
\begin{subequations}
\begin{eqnarray}
K &=& \xi\e^{-\rho} \,,
\\
H &=& \bigg[1+\frac{k}{m}\bigg(\frac{1-\xi^2 \e^{-2\rho}}{\rho} - 2 \xi^2 \Ei(-2\rho)\bigg)\bigg]^{-\frac{1}{k}} \,.
\hspace{10mm}
\end{eqnarray}
\end{subequations}

The function $F(\lambda)$ follows as
\begin{eqnarray}
F(\lambda) &=& \bigg(1+\frac{\lambda}{M}\bigg)^{-N} \,,
\end{eqnarray}
where
\begin{equation}
N \ \equiv \ \frac{n}{k} \,,
\hspace{10mm}
M \ \equiv \ \frac{m}{k} \,,
\end{equation}
so that the energy density is
\begin{eqnarray}
\frac{\mathcal{E}}{v^4g^2}
&=&
%\frac{1}{(1+\frac{\lambda}{M})^{N}} \bigg[
%2 \frac{\xi^2\e^{-2\rho}}{\rho^2}
%+ \frac{N}{M}\frac{(1-\xi^2\e^{-2\rho})^2}{(1+\frac{\lambda}{M})\rho^4}
%\bigg]
%\\ &=&
\frac{M^N}{(M+\lambda)^{N}} \bigg[
2 \frac{\xi^2\e^{-2\rho}}{\rho^2}
+ N\frac{(1-\xi^2\e^{-2\rho})^2}{(M+\lambda)\rho^4}
\bigg]
\,.
\hspace{8mm}
\end{eqnarray}
For $\rho \to 0$ it goes like
\begin{eqnarray}
\frac{\mathcal{E}}{v^4g^2}
&=&
\begin{cases}
\displaystyle
\frac{N M^{N}}{(1-\xi^2)^{N-1}} \rho^{N-3} \big[
1
%\nonumber \\ && {}
%- \frac{
%MN(1+N)
%-2 \xi^2 \big[
%1+2N
%+ N (1+N) \log(2\e^{\gamma_{\mathrm{E}}-1}\rho)\big]
%}{(1-\xi^2) N} \rho
%%\nonumber \\ && {}
%%+ \frac{
%%(...) + (...) \log(2\e^{\gamma_{\mathrm{E}}-1}\rho) + 4 N \xi^4 (1+N)(2+N) \log^2(2\e^{\gamma_{\mathrm{E}}-1}\rho)
%%}{2 (1-\xi^2)^2 N} \rho^2
%\nonumber \\ && {}
+ \mathcal{O}(\rho)
\big]
\,,
& (\xi^2 < 1)
\\[1em]
\displaystyle
\frac{2}{\rho^2} \bigg(\frac{M}{-2\log\rho}\bigg)^N + \ldots
\,,
& (\xi^2 = 1)
\end{cases}
\end{eqnarray}
so it must be
\begin{equation}
N \ \geq \ 3
\hspace{6mm}
\mbox{and}
\hspace{6mm}
\xi^2 \ < \ 1
\,,
\end{equation}
so that the energy density is finite. In contrast to the previous case $k=0$, now the value $\xi^2 = 1$ is not allowed, as that would lead to a divergent $\mathcal{E}$ for any value of $N$.

\section{Summary and discussion}
\label{conclusions}

We have introduced a class of $\group{SU}{2}{}$ non-Abelian gauge theories with the most general structure of kinetic terms involving arbitrary form-functions of the adjoint scalars. We commented on the redundancy of this description due to the field redefinition of the scalar fields. We have focused on establishing a BPS limit and spherically symmetric topological solutions with special emphasis on the regularity of the energy density everywhere. 

Compared to our previous work \cite{Benes:2023nsr}, the novelty of this paper is an identification of a special class of models, in which a magnetic monopole possess a new parameter, $\xi$, that modifies the shape of the energy density, but does not change the total energy in the BPS limit. This new parameter arises due to the special structure of the BPS equations as a constant of integration that is not fixed by any regularity condition.

What remains to be clarified is the physical interpretation of the parameter $\xi$. As of now, the most plausible hypothesis seems to be the presence of a new kind of continuous symmetry, for which the parameter $\xi$ is a zero-mode. One may think that a source of this symmetry could be the presence of the \qm{new} kinetic terms \eqref{termsKinNew}. However, as one example in Sec.~\ref{subsec:power} illustrates, they are not strictly necessary to obtain monopole solutions containing $\xi$. Of course, we do not expect this symmetry to arise from some simple geometrical considerations (as is the case for other well-known zero modes, e.g., the translational zero mode, etc.) and it is perhaps necessary to employ some concept of a generalized symmetry. 

Let us offer a plausible speculation about the $\xi$ moduli that is closely connected to the geometry of the base space. Since $\xi$ arises as a constant of integration of the BPS equation \eqref{logK}, which would normally be fixed by the regularity condition at $\rho=0$, its origin is inextricably linked with the absence of said regularity conditions.

The regularity conditions for the fields at $\rho=0$ are due to a (coordinate) singularity of the spherical coordinates, which are ill-defined at the origin. In our case, we show that we do not need to impose any constraints on $H(0)$ and $K(0)$ to have regular energy density. Furthermore, both the gauge fields and the adjoint scalars are non-analytic at $\rho=0$ for any $\xi$. Specifically, there is a pole for gauge fields and  essential singularity for scalar fields (although $H(\rho)$ and $K(\rho)$ are regular on the positive real axis $\rho \geq 0$). Therefore, it seems to us that the new parameter arises due to the natural extension of the solution beyond the Euclidean space $\mathbb{R}^3$. 

Indeed, we can extend the solutions as $H(|\rho|)$ and $K(|\rho|)$ so that the energy density is \emph{regular on the full line} $\rho \in \mathbb{R}$. This means that our base space is actually a double-sheeted Euclidean space that is joined at $\rho=0$. This space is nothing but a collapsed Ellis wormhole (with the throat width taken to be $0$). There is a physical singularity at $\rho=0$, which explains why our solutions cannot be analytic there.

In other words, we speculate that the $\xi$ parameter is an \qm{echo} of the collapsed wormhole. At this point, however, it is not clear whether $\xi$ has a geometric interpretation in the extended base space or not. We plan to investigate fully analytic solutions in regular wormhole spacetime backgrounds and see what happens to them as we take the wormhole's throat to zero. This should offer us a natural explanation of the $\xi$ parameter. 

Furthermore, there seems to be a natural connection between wormholes and magnetic monopoles, especially in the vanishing throat limit. Indeed, any magnetic flux that passes through the throat that is small enough to be undetectable would be perceived by observes at asymptotic infinity $(\rho \to +\infty)$ as a magnetic monopole and as an anti-monopole at the other side $(\rho \to -\infty)$, even though there is no source of magnetic monopoles in the entire spacetime. From this point of view, it is natural to study monopole-like solutions in the wormhole background, which we plan to do.

To clarify the r\^{o}le of the $\xi$ parameter further, we should study monopole solutions away from the BPS limit, where we expect that $\xi$ will cease to be a free parameter, but would attain a specific, model-dependent value that minimizes the energy functional. 

We plan to elaborate on these observations in the future.

\begin{acknowledgements}
The authors would like to express the gratitude for the institutional support of the Institute of Experimental and Applied Physics, Czech Technical University in Prague (P.~B.~and F.~B.), and of the Research Centre for Theoretical Physics and Astrophysics, Institute of Physics, Silesian University in Opava (F.~B.). P.~B.~is indebted to A\v{s}tar \v{S}eran and FSM for invaluable discussions. The work of F.~B.~is supported by SGS/24/2024 Astrophysical processes in strong gravitational and electromagnetic fields of compact object.
\end{acknowledgements}

\appendix

%\onecolumngrid
%\twocolumngrid

%\bibliographystyle{JHEP}
%\bibliography{references}

\providecommand{\href}[2]{#2}\begingroup\raggedright\endgroup

\end{document}